\documentstyle[aps,prl,twocolumn,epsf,graphicx]{revtex}
\begin{document}
\draft
\preprint{}
\twocolumn[\hsize\textwidth\columnwidth\hsize\csname @twocolumnfalse\endcsname
\title{A Model for Quantum Stochastic Absorption in  Absorbing Disordered Media }
\author{ Prabhakar  Pradhan\cite{ppemail}}
\address{Department of Electrical and Computer  Engineering,
         Northwestern University,  Evanston, IL 60208, USA}   
\maketitle
\date{\today}
\begin{abstract}
Wave propagation  in  coherently absorbing disordered media is generally modeled 
by adding a complex part to the real part of the potential.  In such a case, it is 
already understood  that the complex potential plays a duel role; it acts as an absorber 
as well as a reflector due to the mismatch of the phase of the real and complex parts of 
the  potential. Although this model gives expected results for weakly absorbing 
disordered media,  it gives unphysical results  for the strong absorption regime 
where it causes the system to behave like a perfect reflector. 
To overcome this issue, we develop  a model here 
using {\em stochastic absorption} for the modeling
of absorption by $``$fake$"$, or $``$side$"$, channels obviating the
need for a complex potential. This model of stochastic absorption eliminates the 
reflection that is coupled  with the absorption in the complex potential model 
and absorption is proportional to the magnitude of the absorbing parameter. 
Solving the statistics of the reflection coefficient and  its phase for both the models, 
we argue that stochastic absorption is a potentially better way of modeling absorbing 
disordered media. 
\end{abstract}
\vspace{.0cm}]
\section{introduction}
Quantum and classical wave propagation and localization in 
disordered media is a well studied problem  
\cite{Anderson58,Lee85,Ping90}.   
Recently in the literature, more attention has been paid  
to  studying  the effect of localization on coherent 
absorption/amplification in coherently absorbing/amplifying 
disordered media, both theoretically
\cite{Pradhan94,Pradhan98,Zhang95,Gupta95,Beenakker96,Joshi97,Heinrichs97,Jiang00}
and  experimentally 
\cite{Lawandy94,Ganack94,Lawandy95,Wiersma95a,Wiersma95b,Prasad97,Cao99}.
Light is Bosonic, so it can be  amplified or absorbed. 
On the other hand, electrons are Fermions, which can 
not be amplified; however, they can be absorbed in the sense of 
phase de-coherence. Coherent absorption/amplification  in a 
coherently absorbing/amplifying medium is a non-conserving 
scattering process where temporal phase coherence of a wave 
is preserved despite absorption/amplification. Coherent back 
scattering (CBS) in a disordered medium, the main 
cause of weak and strong localization, is not  affected by 
the additional presence of a coherently absorbing/amplifying  medium due to 
the  persistence of phase coherence of the interfering 
waves. It was  shown \cite{Pradhan94} that  the localization can enhance the 
coherent light amplification  in a coherently amplifying (i.e. lasing) 
disordered medium, resulting in a self-sustaining mirror-less lasing 
action where coherent feed back for lasing action
is  supported by disorder induced localization.
In that same frame work, we  also studied the effect of 
localization on the coherently absorbing disordered media 
and showed the similarities of the two processes for effective 
localization  and effective active lengths. This problem  studied 
further in detail by different  approaches 
\cite{Zhang95,Gupta95,Beenakker96,Joshi97,Heinrichs97,Jiang00}.
The general approach to model coherent amplification and absorption 
is to add a constant complex part  to the real part of the 
potential. Depending on the sign of the complex potential, 
the model shows amplification or absorption.  Modeling 
absorption/amplification by a complex potential, however, 
always gives a reflection part due to the mismatch between 
the real and the imaginary potentials. Studying a delta 
scatterer,  which has  both real and  complex parts, it 
was shown \cite{Rubio93} that  absorption is not a monotonic function of  
the strength of the complex potential.  Although   
complex potential  model  works well for the case of  weakly  
absorbing/amplifying  disordered media, it
gives unphysical results in the strong  
absorption/amplification regime because the system behaves 
like a perfect reflector. 
To address this issue, it  was derived and demonstrated 
\cite{Pradhan98} that absorption without 
reflection can be modeled by using a {\em 
stochastic absorption} model. Recent detailed numerical studies  
have been performed by transfer matrix methods
\cite{Joshi00a,Joshi00b}
to study our approach, and the numerical results  support our model.
Here, we first study the coherently absorbing disordered media 
modeled by a adding complex potential. 
Then, we present a detailed derivation and analysis of the  
Langevin equation in case of  stochastic absorption for the modeling 
of absorption by $``$fake$"$, or $``$side$"$, channels obviating the 
need for a complex potential. We show that  stochastic absorption model  
gives potentially better  physical results  than the  
model by a complex potential for transport in 1D 
absorbing  disordered  media in the regime of strong absorption.
However, these two models give the  same result for weak
disorder and the weakly absorbing regime.
\section{ Coherent Absorption Modeled by  Complex Potentil}
There has been recent interest in the role of absorption
in localization, and on the different length scales in the
problem in the presence of absorption. For the bosonic case,
like light etc., a coherent state (e.g. a laser beam) is an
eigenstate of the annihilation operator. Removal of a
photon(absorption) does not destroy the phase coherence.
Fermions cannot be annihilated in the context we are considering
here. For the Fermionic case, the physical picturization of the
absorption will be some kind of inelastic process (like
scattering by phonons), where electrons lose their  partial temporal
phase memory. This type of absorption is called {\em stochastic absorption}. 

 Recent theoretical studies  \cite{Weaver93,Yusofin94,Jayan94} have
shown that the absorption in the case of light waves does not give
any cut-off length scale for the localization problem. 
If a system is in the localized state, absorption will not kill 
the localization to make the system again diffusive. A sharp
mobility edge exists even in the presence of significant absorption
in 3D.

Adding a constant imaginary part with the proper sign to the real
potential of the Maxwell/Schr\"odinger wave equation can model the linear
coherent  absorption/amplification. Both the Maxwell and the 
Schr\"{o}dinger equations can be transformed to the form of
the Helmholtz equation. A general form of a Helmholtz equation 
can be written as:
\begin{equation}
\frac{\partial^2 u}{\partial x^2} +  k^2 [1+(\eta(x)+i\eta_a)] u\, = 0 \quad,
\label{heq}
\end{equation}
For the electronic case, consider  $V(x)+iV_a$ as the potential, $k$ as the wave vector,
and  $\hbar/2m =1$. For  the optical  case, consider  $\epsilon_0+i\epsilon_a$ as 
the constant dielectric background  and $\epsilon(x)$ as the randomly 
spatially fluctuating part of the dielectric constant, 
$k\equiv 2\pi/\lambda$, where $\lambda =$
wavelength in the average medium $(\epsilon_0)$.
Then, we can  define  $\eta(x)$=$-V(x)/k^2$,
$\eta_a$=$-V_a/k^2$ for the Schr\"{o}dinger equation, and
$\eta(x)$= $+{\epsilon(x)\over \epsilon_0 }$,
$\eta_a$= $+{\epsilon_a\over \epsilon_0}$ 
and $k^2={\omega^2\over c^2}\epsilon_0$ for the Maxwell equation.

The model we are considering here is for a one-channel scalar wave
where the polarization aspects have been ignored. This holds well 
for a single-mode polarization maintaining optical fiber. For electronic
case, we  consider only a single  channel.

As pointed out in Ref  \cite{Rubio93}, modeling absorption 
by a complex potential will always give a concomitant reflection. 
Absorption is not possible without reflection and there is a competition 
between the absorption and the reflection. For  very high absorption, 
the system may try to act as a perfect reflector. We extend and explore 
this idea for coherently absorbing 1-D disordered media. 
The Langevin equation for the complex amplitude reflection
coefficient $R(L)$ can be  derived by the invariant imbedding
method \cite{Rammal87} from the Helmholtz equation Eq.(\ref{heq}),
\begin{equation}
  \frac{d R(L)}{d L} \, = \,\,  
2ik R(L)\, + \,i \frac{k}{2} (1+\eta (L) + i \eta_a )(1+R(L))^2 \quad,
\label{leq}
\end{equation}
with the initial condition R(L)=0 for L=0.

Now, taking $R(L)= \sqrt{r} e^{i\theta }$, the Langevin
equation Eq.(\ref{leq}) reduces to two coupled differential
equations in $r$ and $\theta$.
For a Gaussian white noise  potential, using a stochastic Liouville equation 
for the evolution of probability density and then integrating out  the 
stochastic part by Novikov's theorem, we get the  Fokker-Planck equation: 
\begin{eqnarray}
\frac{ \partial P(r,\theta) }{\partial l} &=& 
  \left[ \sin\theta\frac{\partial}{\partial r}( r^{1/2} (1-r))
 + \frac{\partial }{\partial \theta}  \right. \nonumber \\
&& \left.+ {1\over 2} (r^{1/2} + r^{-1/2}) \frac{\partial}{\partial \theta} 
    cos(\theta) \right ]^2 P(r,\theta)  \nonumber \\
&& - 2 k\xi \frac{\partial}{\partial \theta} 
\left[ P(r,\theta ) \right]  \nonumber \\
&& + D  \frac{\partial}{\partial r} 
\left[ r P(r,\theta ) \right]  \nonumber \\
&& + {D\over 2} \cos \theta \frac{\partial }{\partial r} 
\left[ r^{1/2}(1+r) P(r,\theta ) \right]  \nonumber \\
&& + {D\over 4}  (r^{1/2}-r^{-1/2})
\frac{\partial}{\partial \theta } \left[\sin \theta P(r,\theta ) \right]
\quad, 
\label{fpeq}
\end{eqnarray}
where we have introduced the dimensionless length $l= \frac{L}{\xi}$  
and $D=\frac{4\eta_a}{qk}$ = $\xi/\xi_a$, and $\xi\equiv ({1\over
2}qk^2)^{-1}$ is the localization length and $\xi_a={1\over
2k\eta_a}$ is the absorbing length scale in
the problem, and  $q$ is the strength of the delta correlation 
of the random potential. 

The Fokker-Planck equation Eq.(\ref{fpeq}) can be solved analytically
for the asymptotic limit of large lengths in the random phase
approximation (i.e., for the weak disorder case).
Then, the statistics $P^D(r,l)$ have a steady state 
distribution for $l\rightarrow \infty$. It
can be obtained by setting $ \partial P(r,l)/\partial l = 0 $
and solving the resulting equation analytically. We get \cite{Pradhan94}
the following steady state distribution:
\begin{eqnarray}
\lim_{l\rightarrow\infty\atop }
  P^D(r,l) & = & \left\{\begin{array}{cl}
\frac{|D| exp(|D|) exp(-{|D|\over 1-r})}{(1-r)^2} &
 \quad \mbox{for}\; r\leq 1\\
0 &\quad\mbox{for}\; r > 1,\end{array}\right.
\label{steady}
\end{eqnarray}
For $D=0$ and $l\rightarrow \infty$, equation (\ref{fpeq})  
reduces to a half delta function peaked around $r=1$.
The Fokker-Planck equation, Eq.(\ref{fpeq}), is difficult to
solve analytically beyond the random phase approximation.
We  solve the full Fokker-Planck equation (\ref{fpeq}) numerically
to calculate the statistics of the reflection coefficient and its phase 
in a regime of strong disorder with strong absorption. 

 To see the actual evaluation of  the probability as a function of
reflection coefficent for a fixed disorder strength and length, we 
evolved the full Fokker-Planck equation without approximation.
The Fokker-Planck equation is  solved as an initial value problem,
i.e.  $P(r,l=0)=\delta(r=0)$.

\begin{figure}
\epsfxsize=8cm
\epsfysize=11.5cm
\centerline{\epsfbox{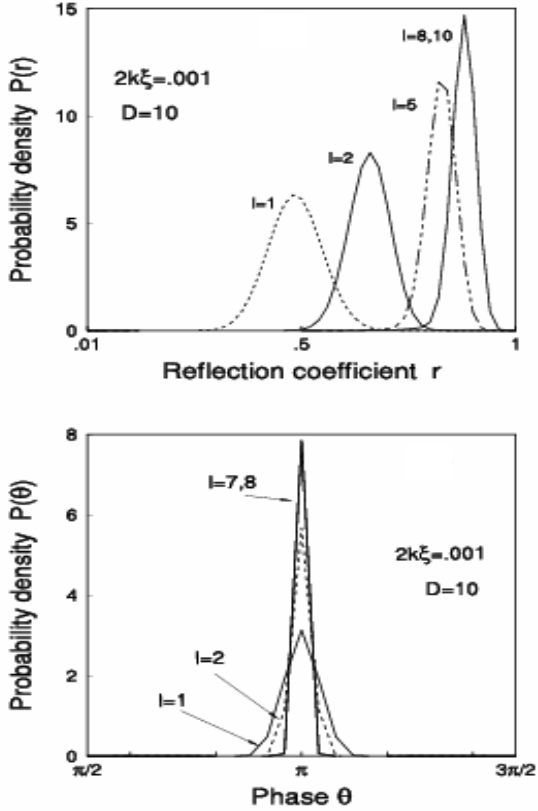}}
\caption{ Probability distribution $P(r)$ and $P(\theta)$ separately
against the sample length  $l$. 
Parameters are fixed strong disorder strength $2k_F\xi
=.001$ and fixed strong absorbing  parameter $|D|= 10$.}
\label{fig12}
\end{figure}

In Fig.\ref{fig12}  we plot  the results of our numerical 
simulations. The $P(r)$ distributions  show (top plot) that for 
the strongly absorbing  disordered media, the probability 
initially peaks near $r=0$ and the peak slowly moves towards 
$r=1$ for larger sample lengths, indicating stronger
reflection for stronger absorption. In Fig.1, the bottom  plot shows that 
the steady state phase distribution is a delta function peaking 
symmetrically about $\theta=\pi$. 

These plots, show  the system behaving  as a perfect reflector 
for higher lengths, with strong disorder and strong
\noindent absorbing  parameter $|D|$. Therefore, this coherent absorbing model  
is restricted  for the weakly  absorbing parameter regime.
\section{ Stochastic Absorption}
 In the previous section, we have shown and discussed that the 
absorption is not possible without reflection 
in a  calculation where absorption 
is modeled by adding a constant imaginary potential to the real 
potential. In the Langevin equation for $R(L)$, derived from 
these types of models, the wave always gets a reflected 
part along with the absorption. There is, however, another 
way of deriving a Langevin equation for the reflection amplitude 
$R(L)$ such that the absorption does not have a concomitant 
reflected part. This approach is motivated by the work of references 
\cite{Buttiker86,Maschke95} where some purely absorptive $``$fake$"$ (side) 
channels are added to the purely elastic scattering channels of 
interest. A particle that once enters to the absorbing channel, 
never comes back and it is physically lost. We
derive the Langevin equation for $R(L)$ following the
approach in Ref.\cite{Buttiker86}. This Langevin equation has some formal
differences from  Eq.(\ref{leq}).
However,  
\begin{figure}
\epsfxsize=7.5cm
\centerline{\epsfbox{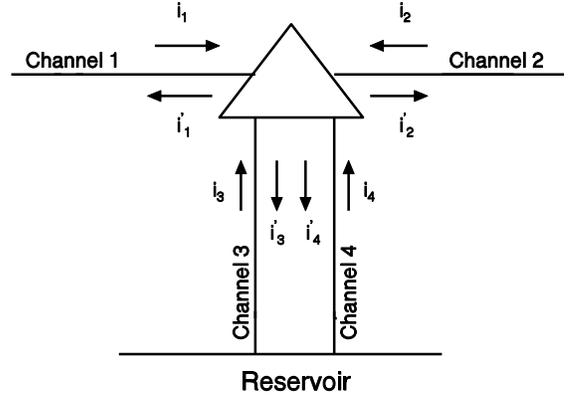}}
\caption{ Modeling $``$absorption$"$ by fake channels:
Channels 1 and 2 are coupled through the  current leads to
two "fake" channels 3 and 4 which connect to a thermal reservoir.}
\label{figfake}
\end{figure}
\noindent it does not differ for the weak disorder case.

Now we proceed to derive Langevin equation for $R(L)$ using the $``$fake$"$ 
channel approach. The main idea is to simulate absorption by enlarging the
S-matrix to include some $``$fake$"$, or $``$side$"$ channels that remove
some  probability flux out of reckoning. 
Let us assume that the scatterer has a general scattering channel
and an absorbing channel ending with a phase randomizer which acts as a blackbody.
The scattering matrix can be taken as discussed in  reference \cite{Buttiker86}. 
Consider  scattering channels 1 and 2 connected through
current leads to two quantum $``$fake$"$ channels 3 and 4 that carry
electrons to the reservoir, ( with chemical potential $\mu$) as
shown in  Fig.\ref{figfake} (for only one scatterer). This is a
phenomenological way of modeling absorption \cite{Buttiker86,Maschke95}.
Electrons entering into the channels 3 and 4 are absorbed regardless
of their phase and energy. The absorption is proportional to the
strength of the $``$coupling parameter$"$ $\epsilon$. 
A symmetric scattering matrix $S$ for such a
system can be written as :
\begin{equation}
\left( \begin{array}{cccc}
            \alpha R  & \alpha T  &  0           & \beta       \\
            \alpha T  & \alpha R  & \beta       &  0          \\
             0        & \beta    & -\alpha R^*  & -\alpha T^*  \\
             \beta    &  0        & -\alpha T^*  & -\alpha R^*   
\end{array} \right) \quad ,
\label{meteq}
\end{equation}
where $R$ and $T$ are the reflection and the transmission
amplitudes for the single scatterer , $\alpha =\sqrt{1-\epsilon}$
is the absorption coefficient, and $\beta=\sqrt{\epsilon}$. 

Now, the full scattering matrix is unitary for all positive real  
$\epsilon < 1$. But the sub-matrix directly connecting only the  
channels 1 and 2  is not unitary. 
We will now explore this fact in our derivation. We observe from the
above sub-matrix that for every scattering involving channels 1 and 2
only the reflection and the transmission amplitudes get multiplied by
$\alpha$ (the absorption parameter). Now keeping this fact in mind,
we will derive the Langevin equation for the reflection amplitude
$R(L)$ for $N$ scatterers (of length L) each characterized by a random
S-matrix of this type, with the same $\alpha$ value.
\begin{figure}
\epsfxsize=7.6cm
\centerline{\epsfbox{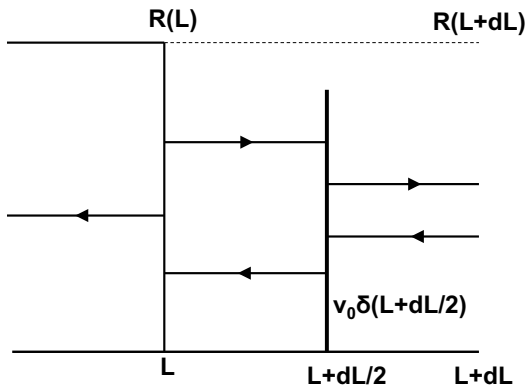}}
\caption{ A schematic picture of a scatterer of length $L$ having reflection 
amplitude $R(L)$ while an added length $dL$ makes the length of the scatterer 
$L+dL$. The effective potential of the length $dL$ is shown by an effective summary 
delta-function potential.}
\label{fig_del}
\end{figure}
 Consider $ N $  random scatterers for a 1D sample of length $L$ 
with $(\equiv Na, a\equiv unit spacing)$ the reflection amplitude
$R(L)$, and $N+1$ scatterers of the sample length $L+\triangle L$,
with the reflection amplitude $R(L+\triangle L)$. That is, let us
start with the $N$ scatterers, and add one more scatterer to the
right to make up the N+1 scatterers. Now, we want to see the relation
between $R(L)$ and $R(L+\triangle L)$. That is, let there be a
delta-function potential scatterer between $L$ and $L+\triangle L$,
positioned at the point $L +\triangle L/2$, that can be considered as
the effective scatterer due to the extra added scatterers for length
$\triangle L$. 
For $k\triangle L\ll 1$, we can treat the extra added scatterer as an
effective delta potential $v_0(L) \delta (x-L- \triangle L/2)$
with $v_0(L) = V(L) \triangle L $.(We consider the continuum limit, $a\rightarrow 0$,
 $N\rightarrow \infty$, and $Na=L$ fixed ).

Now, for a plane-wave scattering problem for a 
delta function potential of strength $v_0$, which is at $x=0$
and has complex reflection and transmission amplitudes $r_0$ and $ t_0$
respectively, we have from the continuity condition for the wave function
and discontinuity condition for the derivative of the wave
function (which one gets by integrating the Schr\"{o}dinger equation
across the delta function):
\begin{eqnarray}
 r_0 = \frac{v_0}{i 2 k - v_0}\quad \mbox{and} \quad 
 t_0 = r_0+1. \nonumber
\end{eqnarray}
Considering $ v_0 = v(L) \triangle L $ the smallness parameter, one gets
expressions up to first order in $\triangle L$ for $r_0$ and $t_0$:
\begin{eqnarray}
r_0 = \frac{v_0 }{2ik} \quad  \mbox{and}\quad
t_0 = 1 + \frac{v_0 }{2ik}, \nonumber
\end{eqnarray}
where we have taken $\hbar^2/2m=1$.

Now to introduce absorption we  will write:
\begin{eqnarray}
 r &&\rightarrow r \alpha = r_0(1-\epsilon)^{1/2} , \nonumber\\ 
  t &&\rightarrow t \alpha = t_0(1-\epsilon)^{1/2} . \nonumber
\end{eqnarray}
This means that for every scattering the reflection and transmission
amplitudes get modulated by a factor of ${\alpha}$.

Now consider a plane wave incident on the right side of the
sample of length $L+\triangle L$. Summing all the processes of
direct and multiple reflections and transmissions, on the right
side of the sample of length $L$, with the effective delta
potential at $L+\triangle L/2$, we get, 
\begin{eqnarray}
 R(L+\triangle L) &=& r e^{ik\triangle L} \nonumber \\
    &+& e^{ik\triangle L/2} t e^{ik\triangle L/2}  R(L) 
           e^{ik\triangle L/2} t e^{ik\triangle L/2} \nonumber \\
       &+& ....... \,.
\label{leqst}
\end{eqnarray}
 Summing the above  geometric series, substituting the values of $r$ and $t$, 
and taking the continuum limit for L , one gets from Eq.(\ref{leqst}),
the Langevin equation:
\begin{equation}
 \frac{d R(L)}{d L} \, = \, - \alpha R(L) +  
2ik R(L)\, + \,i \frac{k}{2} (\eta(L))(1+R(L))^2 \, ,
\end{equation}
with the initial condition $R(L)=0$ for $L=0$, and $\alpha $ is the
absorbing parameter. 
This is not quite the same as the Eq.(2) obtained by
introducing an imaginary potential. It turns out, however, that 
this Langevin equation gives the same results in the regime of weak 
disorder but differs qualitatively in the regime of strongly 
absorbing disordered media.

Similarly, following the same steps as we discussed previously,
from the Langevin Eq.(\ref{leqst}) we derive the Fokker-Planck
equation: 
\begin{eqnarray}
\frac{\partial P(r,\theta)}{\partial l} &=&
 \left[ \sin \theta \frac{d}{dr}( r^{1/2} (1-r))
       +   \frac{\partial}{\partial \theta} \right. \nonumber \\
 &+& \left. {1\over 2} (r^{1/2} + r^{1/2}) \frac{\partial}{\partial \theta} 
     (  cos(\theta))\right] ^2 P(r,\theta) \nonumber \\
 &-&2 k\xi \frac{\partial P(r,\theta)}{\partial \theta}
 + D  \frac{\partial(Pr)}{\partial r} \quad ,
\label{fpst}
\end{eqnarray}
where parameters $l$ and $2k\xi$ are same as defined in Eq.(\ref{fpeq}),
and $D={4\alpha\over qk}$.

In the weak disorder limit,
when the phase part is integrated out by the RPA, 
Eq. (\ref{fpst}) and Eq.(\ref{fpeq}) give the same Fokker-Planck equation in $r$.
Hence, the steady state $P(r)$  solution for the  both are same, i.e.   
Eq.(\ref{steady}).
We will consider here only the strong disorder case for
the numerical calculations of the Fokker-Planck equation
(\ref{fpst}), which goes beyond the RPA to compare it with
the solution of Eq.(\ref{fpeq}). 

\begin{figure}
\epsfxsize=8.5cm
\epsfysize=11.5cm
\centerline{\epsfbox{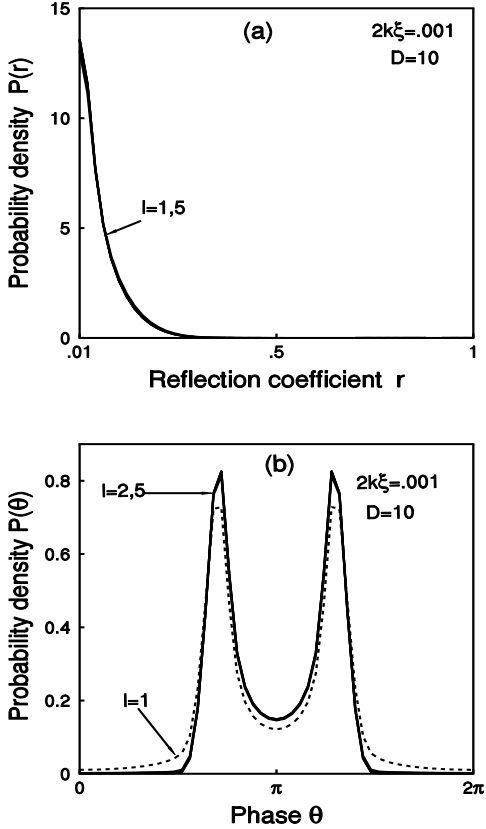}}
\caption{ Probability distributions (a) $P(r)$ and (b) $P(\theta)$
against the sample length $l$ with
fixed strong disorder strength $2k_F\xi =.001$ and
fixed strong stochastic absorption parameter $D= 10$.}
\label{fig14}
\end{figure}

Fig.\ref{fig14}(a) is the plot of $P(r)$ for the strong disorder ($2k_F\xi=.001$)
and strongly absorbing ($D=10$) limit, with  different lengths of the sample. 
$P(r)$ distributions are  peaked around $r=0$. The wave is absorbed in the 
medium before it gets reflected as we have
modeled absorption without reflections and we are considering
the case of strong absorption, and hence the probability of
absorption is more than that of reflection. 
Fig.\ref{fig14}(b) shows the phase distribution $P(\theta)$,
which is a double peaked symmetric distribution and differs from
the model of absorption by complex  potential, which is  a
delta function distribution at $\theta=\pi$ as shown in Fig.1 (bottom).
\section{ Discussion and Conclusion}
In conclusion, we have shown a comparison between the modeling of 
absorption in disordered medium (i) by a complex potential and (ii) by
fake channel $S$  matrix approach.
We derived  a new  Langevin equation using the latter method. 
We calculated the statistics of the reflection
coefficient $(r)$ and its associated phase $(\theta)$ for a wave
reflected from absorbing  disordered media, for
different disorder strengths and lengths of the sample, for the both models.
For the case of weak disorder with weak absorption, i.e. within the  RPA, 
the FP equation for the case of stochastic absorption is the same as 
that of the  absorption modeled by a complex potential. For the regime
of strong disorder with strong absorption, the  {\em stochastic absorption} model
behaves as a perfect absorber opposite to the coherent absorbtion case,
which behaves as a perfect reflector.  
Therefore, the Langenvin equation derived from stochastic absorption method 
gives a  potentially better physical model for coherent absorption. 
Our  model proposed here has  already been  verified and supported  by extensive 
numerical simulation  using a transfer matrix method \cite{Joshi00a,Joshi00b}.   

I gratefully acknowledge N. Kumar for many stimulating discussions.

\end{document}